\documentclass{WileyMSP-template}
\usepackage{color}
\usepackage[colorlinks,linkcolor=red]{hyperref}
\begin{document}

\pagestyle{fancy}
\rhead{\includegraphics[width=2.5cm]{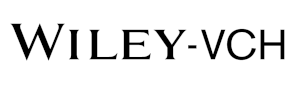}}

\title{Nanoscale positioning approaches for integrating single solid-state epitaxial quantum emitters with photonic nanostructures}

\maketitle

\author{Shunfa Liu}
\author{Kartik Srinivasan*}
\author{Jin Liu*}



\begin{affiliations}
	Dr. Shunfa Liu \\
	State Key Laboratory of Optoelectronic Materials and Technologies, School of Physics, Sun Yat-sen University, Guangzhou 510275, China\\

	Dr. Kartik Srinivasan\\
	Microsystems and Nanotechnology Division, Physical Measurement Laboratory, National Institute of Standards and Technology, Gaithersburg, Maryland 20899, USA\\
	Joint Quantum Institute, NIST/University of Maryland, College Park, Maryland 20742, USA\\
	Email Address:kartik.srinivasan@nist.gov
	
	Prof. Jin Liu \\
	State Key Laboratory of Optoelectronic Materials and Technologies, School of Physics, Sun Yat-sen University, Guangzhou 510275, China\\
	Email Address:liujin23@mail.sysu.edu.cn
	
\end{affiliations}



\begin{abstract}

Deterministically integrating single solid-state quantum emitters with photonic nanostructures serves as a key enabling resource in the context of photonic quantum technology. Due to the random spatial location of many widely-used solid-state quantum emitters, a number of positoning approaches for locating the quantum emitters before nanofabrication have been explored in the last decade. Here, we review the working principles of several nanoscale positioning methods and the most recent progress in this field, covering techniques including atomic force microscopy, scanning electron microscopy, confocal microscopy with \textit{in situ} lithography, and wide-field fluorescence imaging. A selection of representative device demonstrations with high-performance is presented, including high-quality single-photon sources, bright entangled-photon pairs, strongly-coupled cavity QED systems, and other emerging applications. The challenges in applying positioning techniques to different material systems and opportunities for using these approaches for realizing large-scale quantum photonic devices are discussed.

\end{abstract}


\section{Introduction}
Solid-state quantum emitters are playing increasingly important roles in building quantum networks and realizing on-chip quantum information processing tasks~\cite{Aharonovich2016,1lodahl2015interfacing}. Their discrete energy levels can be exploited to trigger the generation of quantum states of light, including single photon Fock states and entangled photon pairs, enabling the realization of on-demand quantum light sources whose performance can exceed that of counterparts based on non-linear optical processes~\cite{Senellart2017} (e.g., spontaneous parametric down conversion). In addition, the electron and nuclear spins associated with certain types of solid-state quantum emitters (e.g., color centers in diamond or hole spins in quantum dots) can exhibit long coherence times, allowing the high-fidelity initialization, ultra-fast manipulation, and efficient readout of the spin states~\cite{Atature2018}. To fully explore the potential of solid-state quantum emitters in the context of quantum photonic technology, e.g., building highly-efficient quantum light sources and quantum logic gates, photonic naonstructures are inevitably introduced for enhancing the efficiency of light extraction and the strength of light-matter interaction simultaneously.
The last couple of decades has witnessed the tremendous success of novel quantum photonic devices based on such coupling between single solid-state quantum emitters and highly-confined optical modes provided by photonic nanostructures~\cite{1lodahl2015interfacing}. However, many types of high-performance solid-state quantum emitters are randomly distributed within the plane of the host material, which imposes a great challenge in realizing optimal spatial overlap between the quantum emitter and the targeted optical mode supported by the photonic nanostructure. In addition, solid-state quantum emitters are invariably inhomogeneously broadened, due to variations in their size, shape, composition, or proximal environment, at a level that is usually much wider than the operation bandwidth of the photonic nanostructures (particularly the case for high quality factor microresonators, for example). This inhomogeneous broadening hinders the ability to spectrally match a given emitter to a nanostructure, even if it is appropriately spatially located. In early years, coupled devices were typically realized by fabricating a large number of devices and looking for suitable candidates one at a time, with the fraction of devices showing appreciable coupling typically at the 1~\% (or much lower) level. This approach enabled a number of demonstrations of physical phenomena~\cite{Yoshle2004,Gayral2000,Santori2003}, but is intrinsically probabilistic and time consuming, making it challenging to further scale up systems to incorporate multiple identical quantum emitters for more advanced experiments and applications, such as quantum simulation\cite{Wang2017,Loredo2017}. Therefore, in recent years numerous techniques have been developed for controlling the spatial and spectral position of quantum emitters with respect to photonic nanostructures, thereby enabling deterministic coupling to be achieved.

\indent In this review, we focus on nanoscale positioning techniques developed for epitaxial quantum dots (QDs), due to their excellent coherence properties and compatibility with standard nanofabrication processes. There are multiple approaches for nanoscale positioning of epitaxial QDs with respect to targeted photonic structures, including pre-patterning the substrate before the growth to realize ordered QD arrays, heterogeneous integration in which QD-containing structures are removed from their native platform and placed within a photonic environment created in a separate platform, and location of the QDs with respect to fiducial alignment marks after their growth. In the first case, the QDs can be grown at pre-specified locations that are known during subsequent fabrication of photonic nanostructures, providing an approach that is, in principle, more scalable when integrating multiple QD devices. However, the optical properties of these so-called site-controlled QDs usually suffer from the defects in the etched surfaces induced during the substrate patterning process~\cite{Schneider2008,Albert2010}, though it should be noted that promising results with respect to homogeneity (between quantum dots) and single-photon purity and coherence have been reported~\cite{zhang_highly_2019,grose_development_2020}. Heterogenous integration, via pick-and-place techniques of QDs in nanowires, for example, is an exciting and yet technically demanding path for combining the strengths of different materials platforms. Progress along this line was recently summarized in two review articles~\cite{Kim2020a,Elshaari2020} and will not be discussed further. Instead, we aim to cover approaches for which a pre-selected QD with desired optical properties, e.g., suitable wavelength, narrow linewidth and high brightness, etc., is located with nanometer-scale accuracy and precision with respect to alignment marks that are subsequently utilized in photonic nanostructure fabrication. Instead of focusing on the device performance in the context of the deterministically fabricated quantum light sources~\cite{Rodt_2020}, we will focus on the nanoscale positioning techniques themselves, including those based on scanning microscopies, \textit{in situ} lithography, and wide-field fluorescence imaging. For each, we will note the basic working principle, experimental implementations, and strengths and weaknesses. The advantages and challenges for each method will be thoroughly discussed within the context of key quantities including positioning accuracy, speed, and experimental complexity. Finally, we note that while the positioning techniques that we describe can be applied to a wide variety of solid-state quantum emitters~\cite{Aharonovich2016}, including defects in 2D materials~\cite{chakraborty_advances_2019}, color centers in crystals~\cite{atature_material_2018}, and colloidal quantum dots and other nanoparticle emitters~\cite{michler_quantum_2000}. In considering the nanoscale positioning approaches we describe in this review in the context of those systems, it is important to note that they may have other approaches available for realizing precise location information for the quantum emitters - for example, site-controlled defect formation in 2D materials via strain engineering~\cite{branny_deterministic_2017,palacios-berraquero_large-scale_2017}, selective irradiation to form defect color centers in crystals~\cite{schroder_scalable_2017}, and the use of scanning probe microscopy to physically move nanocrystal structures with high accuracy and precision~\cite{schell_scanning_2011}.

\section{Scanning-based positioning techniques}

The epitaxial QDs used in quantum device applications are typically covered by a semiconductor capping layer (wider bandgap than that of the QD material) that is several tens of nanometers in thickness, preventing adverse influences from the environment and thus enabling stable optical properties, including a near-unity radiative efficiency and suppression of blinking effects that are common to many colloidal QDs~\cite{Bayer2019}. However, a negative consequence of this thick capping layer is that it is highly nontrivial to obtain location information of such embedded emitters by surface imaging techniques. In this section, we discuss three approaches to overcome this challenge, based on scanning electron microscopy, atomic force microscopy, and confocal photoluminescence microscopy. The main commonality between these approaches is that 2D images are constructed by raster scanning in a pixel-by-pixel fashion.

\subsection{Scanning electron microscope (SEM)-based positioning approach}

The first deterministically fabricated QD-cavity device was achieved in 2005 by Badolato  \textit{et al.}~\cite{Badolato2005}, and was the result of a joint development between specifically engineered epitaxial growth and SEM imaging. In order to obtain the spatial information from the surface of the sample, 6 layers of vertically strain correlated InAs/GaAs QDs were grown in a way that the QDs were aligned in the vertical direction due to the propagation of the strain, forming a tracer on the sample surface. The surface tracer can be easily detected by SEM for locating the underlying seed QD (only the seed QD is optically active). By depositing alignment marks around the tracer, S1 type (one hole missing in a square lattice) photonic crystal cavities were fabricated around the QD via aligned electron-beam (E-beam) lithography and standard plasma and wet etching to transfer the mask pattern into the semiconductor layer surrounding the QD, as shown in \textbf{Figure  \ref{fig.1}(a)}. This positioning technique brought the targeted QD within $\approx$25 nm of the maximal electric field in the cavity, resulting in the observation of enhancement of emission intensity and radiative decay rate for the QD spectrally resonant with the cavity mode. This QD-cavity system was operating in the weak coupling regime of cavity QED (decay rates exceed the coherent coupling rate) due to the remaining non-negligible spatial mismatch between the QD and the cavity and the relatively low quality factor (Q $\approx$ 3000) of the cavity mode. While the multiple layers of vertical strain-correlated QDs on one hand created surface tracers for positioning the QDs with simple SEM imaging, they also result in greater absorption of light and limited the achievable Qs of the cavity modes. Nearly ten years after the aforementioned publication, Kuruma \textit{et al.} have reported on the ability to use SEM imaging to locate surface features associated with the strain field produced by a buried epitaxial QD without the need for stacked QD layers~\cite{Kuruma2016}. By imaging with a low accelerating voltage (1 kV) and a secondary electron detector placed at a low position with respect to the sample, they detected 1~nm to 2~nm height surface bumps indicative of the underlying QDs, as presented in  \textbf{Figure  \ref{fig.1}(b)}, and conducted an investigation of the coupling with respect to photonic crystal cavity modes as the QD separation from the field maximum varied.

\subsection{Atomic force microscope (AFM)-based positioning approach}

\begin{figure}
	\includegraphics[width=0.95\linewidth]{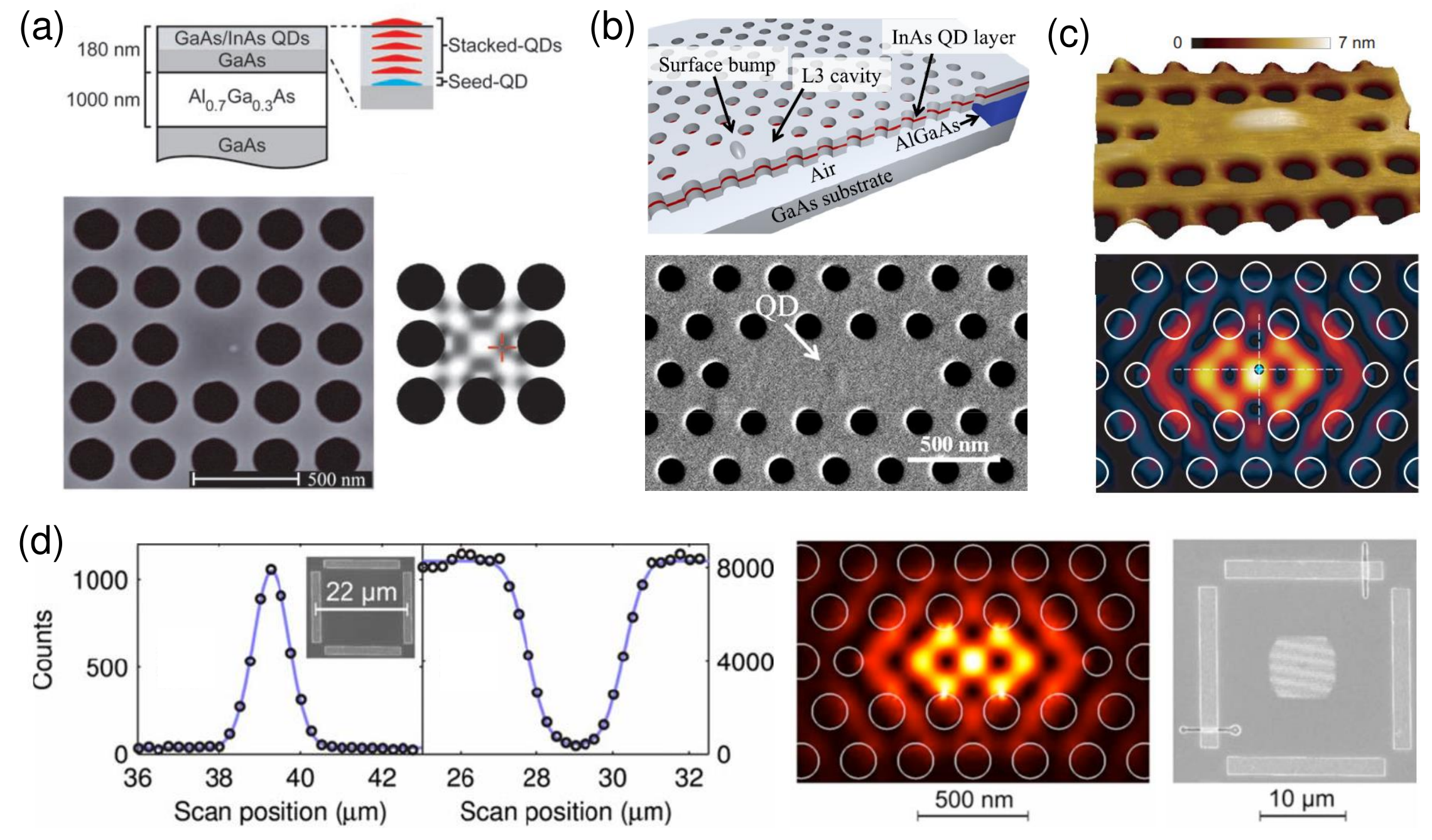}
	\centering
	\caption{Scanning-based positioning technique: (a),(b) Scanning electron microscope (SEM)-based approach used to detected the surface tracer of stacked QDs or 1~nm to 2~nm height surface bumps induced by the underlying QDs. (c) Atomic force microscope (AFM)-based approach employed to trace the tiny surface bumps upon single QD for the realization of strongly coupled QD-cavity system. (d) Confocal PL scanning technique used to probe the location of QDs related to the alignment marks. (a) Reprinted from~\cite{Badolato2005}, with the permission of AAAS. (b) from \cite{Kuruma2016}, with the permission of AIP Publishing (c) \cite{Hennessy2007} Copyright \copyright  2007 Nature Publishing Group. With permission of Springer. (d) Reproduced from \cite{Thon2009}, with the permission of AIP Publishing.}
	\label{fig.1}
\end{figure}

In 2007, Hennessy \textit{et al.}  used an atomic force microscope (AFM) to successfully position single layer QDs embedded in the center of 126 nm thick GaAs membranes without relying on stacked QDs to make surface features observable~\cite{Hennessy2007}. Instead, AFM was employed to trace the tiny surface bumps, as shown in \textbf{Figure  \ref{fig.1}(c)}.  (on the order of 1~nm to 2~nm in height similar to that reported in the subsequent low voltage SEM imaging discussed above). Once the positions of QDs were extracted from the AFM images, similar fabrication methods as those described above were used to create coupled QD-cavity systems. Notably, a strongly coupled QD-cavity system was unequivocally demonstrated via the clear anti-crossing behavior when tuning the QD across the cavity resonance, and photon statistics of the emitted field were also studied. While the positioning accuracy of $\approx$30 nm is similar to the previous work, the strong coupling was achieved due to the significantly improved Q-factor up to the 13300. Such a high-Q factor benefited from the optimized cavity design and the reduced density of QDs by using a single layer configuration. In addition, working with an ultra-low density of QDs within a single QD layer guaranteed the coupling between only one QD and the cavity mode, allowing systematic study of the non-resonant QD cavity feeding with unprecedented clarity~\cite{Winger2009}. As noted in the previous sections, recently it was demonstrated that SEM with optimized scanning/detection parameters can also resolve the surface features associated with sub-surface epitaxial QDs. While maintaining similar positioning accuracy, the QD imaging time was reduced from $\approx$5 min with AFM to several tens of seconds with SEM~\cite{Kuruma2016}. Compared to AFM, SEM is generally more flexible with regards to the magnification, image size and the point of observation, and this much faster observation time suggests that it is more likely to be widely adopted as a surface imaging technique for locating single QDs with respect to photonic nanostructures. However, neither AFM or SEM have been used in as widespread a fashion as the fluorescence based approaches we discuss in the next sections.

\subsection{Confocal micro-photoluminescence ($\mu$-PL) based positioning approach}

A drawback of the positioning techniques based on the identification of surface morphology associated with the underlying QDs is that the geometric centers of the surface features are not necessarily coincident with the centers of the electric dipoles in the QDs~\cite{Sapienza2017}, simply because it is not guaranteed that the strain fields will propagate vertically upward without an angle. In addition, morphological features on surfaces that are correlated with embedded QDs may vary significantly for samples grown by different molecular beam epitaxy(MBE) machines (see Section 6), and even for samples grown in different batches in the same machine. As these methods do not reveal the optical properties of the QDs (in particular, their wavelengths), they must be combined with photoluminescence measurements before subsequent photonic nanostructure fabrication.
Confocal fluorescence microscopy has long been used in the biosciences community to reach nanometer-scale precision in the localization of fluorescent emitters~\cite{Segers-Nolten2002,Betzig2006,Heilemann2008,Thiele2020}. In particular, the position of isolated fluorescing objects can be determined at a level well below the diffraction limit, and this is precisely the scenario of greatest interest to the topic of this paper – namely, the optimized interaction of a single quantum emitter in a nanophotonic structure. In 2009, Thon \textit{et al.} demonstrated how this approach could be used to determine the position of QDs with respect to alignment features to realize strongly coupled cavity QED systems in a deterministic fashion~\cite{Thon2009}, see \textbf{Figure  \ref{fig.1}(d)}. Instead of scanning an electron beam as in SEM or a probe tip as in AFM, in the confocal PL scanning process the sample was moved along the vertical and horizontal directions so that the excitation laser (at 780 nm) traverses both QDs and alignment marks. Two avalanched photodiodes (APDs) were used, one for recording signals related to the QDs and the other used for the alignment marks, with narrowband interference filters placed before the APDs to provide spectral discrimination. The QD emission was monitored near 920 nm, and when the laser is scanned over its spatial location a peak in the fluorescence signal is obtained. Two options for locating the alignment mark positions were available. One was to look at the reflected laser signal at 780 nm, where a peak in intensity would be expected when traversing from the GaAs surface (30~\% reflection) to the metallic surface ($\textgreater$90~\% reflection). Instead, the authors looked for the GaAs band-edge emission near 820 nm, so that a dip in the intensity profile is produced when the laser scans over the metallic alignment mark. By repetitive scanning along the x and y directions, with iterative adjustments to the scan range made to compensate for drifts (in contrast to the biological experiments, epitaxial QD localization via fluorescence requires a cryogenic optical setup), a positioning accuracy as high as ~10 nm was achieved. With this positioning technique, strongly coupled QD-cavity systems with a device yield close to 70~\% were achieved, exhibiting great potential for creating scalable quantum photonic device based on the interactions between multiple QDs. However, the point-by-point nature of the acquisition and multiple iterative scans resulted in localization time scales of about 30 minutes per QD, limiting the overall throughput of this approach.

\section{Wide-field PL imaging-based positioning techniques}

While the scanning-based techniques offer stable access to the spatial information of the single QDs, it in general suffers from the very long times associated with image formation through a line-by-line raster scanning process. Therefore, it is highly desirable to develop techniques that can increase the overall throughput of the process, which is a necessity to build future technologies. As is the case with scanning confocal fluorescence microscopy, the biological sciences again point to a solution, namely the adoption of camera-based imaging techniques which, together with wide-field illumination of the sample and excitation of the embedded quantum emitters, enabled spatially multiplexed measurements to be performed. Such fluorescence imaging approaches have been widely used to track the position of single nanoscale particles in biophysics~\cite{Yildiz2003,Mortensen2010,Smith2010}, and thus offer the possibility of positioning of single quantum emitters in the solid-state. These developments have been enabled by great advances in underlying technologies, for example, single-photon sensitive image sensors such as electron multiplying charge-coupled devices(EMCCD) and scientific complementary metal oxide semiconductor(sCMOS) cameras, and light emitting diodes (LEDs) with enough output power that QD saturation intensities can be achieved across a wide field of view.

\subsection{Single color PL-imaging approach}

The first implementation of PL imaging for achieving deterministic coupling between epitaxial QDs and photonic nanostructures was performed by Kojima \textit{et al.}~\cite{Kojima2013Accurate}, in which strong coupling between single InAs QDs and GaAs photonic crystal cavities was achieved. In their pioneering experiment, metallic alignment marks were first fabricated on semiconductor wafers, in a similar fashion as was done in scanning-based positioning techniques. By illuminating the wafer with a halogen lamp over the whole field of the view of the microscope (see  \textbf{Figure  \ref{fig.2}(a)}), a single optical image with both bright PL spots from QDs and strongly reflecting light from alignment marks were obtained for extracting their relative positions, as shown in  \textbf{Figure  \ref{fig.2}(b)}. The central positions of the bright spots (i.e., the QDs) were then determined by image analysis with sub-pixel size resolution. The positioning accuracy in this work achieved was about 50 nm, as presented in  \textbf{Figure  \ref{fig.2}(c)}, and was largely limited by the signal to noise ratio of the PL images. The major limitation in this experiment is the inability to independently tune the brightness of the QDs and alignment marks in the same image. Optical excitation of QDs with incoherent light generally requires a high excitation intensity, while the metallic alignment marks with strong reflectivity can be imaged with very weak light intensity. Therefore, it turns out to be rather difficult to achieve a single image of both QDs and alignment marks with reasonably good contrast and signal-to-noise ratio if a single illuminator source is used. Several options for remedying this situation are possible. Multiple illumination sources at different wavelengths can be used, so that a single image can be obtained in which the QD emission intensity and alignment mark reflected intensity can be independently controlled. Alternately, multiple images – one of the QDs and one of the alignment marks – can be taken and combined, with some modification to the system (illumination power/wavelength or output spectral filtering) made in-between the images. We discuss both types of approaches below.

\subsection{Bi-chromatic PL-imaging approach}

The aforementioned challenge related to single-color PL imaging for QD location was addressed by Sapienza \textit{et al.} by illuminating the sample with two LEDs with different colors, which we refer to here as a bi-chromatic PL imaging technique~\cite{Sapienza2015}, as schematically shown in  \textbf{Figure  \ref{fig.2}(d)}. The researchers employed two individual LEDs emitting at 630 nm and 940 nm respectively. The full power of 630 nm LED was used to excite the QDs and was subsequently filtered out by a long-pass filter before entering the camera. The wavelength of the other LED for illuminating the alignment marks was selected to be close to the QD emission wavelength in order to reduce the chromatic aberration in the PL images. The brightness of the QDs and the alignment marks in the same image can be independently fine-tuned by controlling the powers of the two LEDs, forming optical images containing both QDs and alignment marks with good signal to noise ratio and high image contrast, as shown in  \textbf{Figure  \ref{fig.2}(e)}. With such improvements in the image quality, the positioning accuracy was improved by a factor of 2 to 3 over that achieved in the single-color positioning technique, as presented in \textbf{Figure  \ref{fig.2}(f)}. The validity of this technique was verified with the realization of bright single-photon sources using QDs deterministically embedded in circular Bragg gratings~\cite{Sapienza2015}.

\subsection{Second Generation of Bi-chromatic PL-imaging approach}

\begin{figure}
	\includegraphics[width=0.9\linewidth]{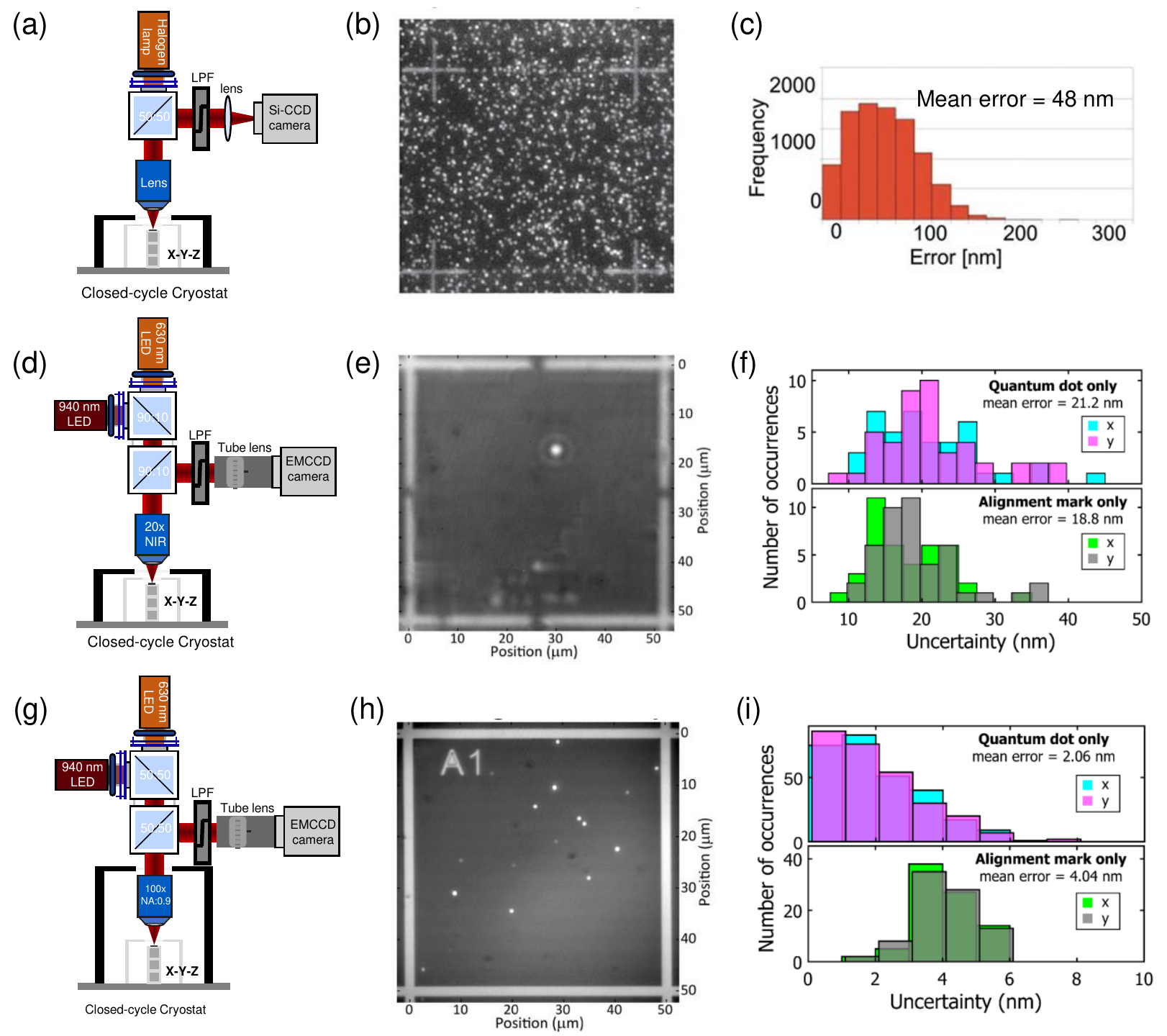}
	\centering
	\caption{Wide-field PL imaging-based positioning techniques. Setup, image, and positioning accuracy of single-color PL-imaging approach (a,b,c), first generation Bi-chromatic PL-imaging approach (d,e,f) and second generation Bi-chromatic PL-imaging approach with objective placed inside the cryostat (g,h,i). (b-c) Reproduced from~\cite{Kojima2013Accurate},  with the permission of AIP Publishing. (e-f) \cite{Sapienza2015} \copyright 2015 Macmillan Publishers Ltd. All rights reserved. With permission of Springer. (h-i) Reprinted with the permission of AIP Publishing from \cite{Liu2017}.}
	\label{fig.2}
\end{figure}

Although the image quality and thus the positioning accuracy are improved via the two-color illumination technique, the overall performance - e.g., positioning time, image quality and the positioning accuracy - can still be further optimized with the improvements both in the hardware and the software (image analysis) of the positioning setup. The accuracy and speed of extracting the center of the QD emission is proportional to the number of photons that are emitted by the QD and subsequently collected by the imaging objective. Therefore, it is highly preferable to employ an objective with a high numerical aperture (NA) for increasing the collected photon flux. In Ref.~\cite{Kojima2013Accurate,Sapienza2015}, the unavoidable distance between the optical window and sample holder of the cryostat prevented higher NA objectives (which have shorter working distances than moderate-to-low NA objectives) from being used, imposing a limit on the achievable photon count rates in the camera. In the second generation of the bi-chromatic PL-imaging system, an objective with NA as high as 0.9 was placed inside the cryostat, which not only ensures that saturation of all QDs within the field of view can be achieved, but also greatly increases the solid angle over which emitted photons by the QDs are collected, as presented in  \textbf{Figure  \ref{fig.2}(g)}. As a consequence, the positioning accuracy was significantly improved (factor of almost 10) and the image acquisition time was reduced from 120 s to 1 s~\cite{Liu2017}. The placement of the objective inside the cryostat offers a few advantages for obtaining high quality PL images, e.g., such a configuration effectively reduces the cryocooler-induced relative vibration between the objective and the sample, and the absence of optical windows between the objective and the sample gives rise to better image quality (in particular, with respect to the alignment marks), as shown in  \textbf{Figure  \ref{fig.2}(h)}. On the image analysis side, two-dimensional maximal likelihood estimation and cross-correlation method replaced the conventional Gaussian fits used in the early work to more reliably extract the positions of the QDs and alignment marks. With the improvements in both hardware and the software in the second generation of the bi-chromatic PL positioning setup, multiple QDs can be positioned by taking a single PL image of the sample with an accuracy as high as 5 nm within only 1 s, as presented in  \textbf{Figure  \ref{fig.2}(i)}. Such short image acquisition times suggest the potential to map full QD wafers within reasonable times (e.g., 1 hr), of importance in scaling up manufacturing of single QD photonic technologies.

\subsection{Further developments and considerations}

Bichromatic excitation is just one approach for constructing images that display both the QD fluorescence and alignment marks. In particular, while this approach relies on the reflection contrast between the metallic alignment marks and adjacent GaAs to enable alignment mark location, the spatially homogeneous emission present in epitaxial QD systems (e.g., due to the GaAs band-edge or the wetting layer in the case of Stranski-Krastanov QDs) provides another approach. Similar to the scanning $\mu$-PL approach of Thon \textit{et al.}, QD imaging experiments have recently been performed in which alignment marks are identified through wetting layer emission, which is essentially uniform in the GaAs regions and dark in the alignment mark regions~\cite{Pregnolato2020}. This image and the QD image are taken sequentially, as the wetting layer emission must be spectrally rejected for the QDs to be observable. The authors report that for their setup, adopting this approach has advantages in comparison to the bichromatic imaging technique.In particular, they have found that imaging the alignment marks by reflecting a second illumination source off the sample surface can result in errors based on the alignment of the illumination source angle. In contrast, identification of the alignment marks by an emission process (i.e., the wetting layer emission) avoids this potential effect. However, it should be noted that previous authors~\cite{Sapienza2015} preferred the bichromatic imaging approach for alignment mark identification because the wetting layer approach requires introduction and removal of an additional filter, which can introduce imaging errors as well. Ultimately, the specific details of the PL imaging setup will be important in assessing the relative errors associated with these different approaches.

\par
Compared to the scanning-based positioning approaches, the PL imaging technique takes advantage of wide-field illumination and multiplexed detection on a sensitive camera to interrogate a large sample area in a relatively short time. In comparison to techniques that require access to separate characterization tools such as SEM and AFM, this approach can be realized through straightforward modification of typical $\mu$-PL setups that are already in place for QD spectroscopy. Through appropriate choice of excitation and/or illumination LED colors, the approach can be adapted for QDs emitting at different wavelengths, such as droplet QDs emitting at 780 nm~\cite{Liu2019}, or to accommodate different epitaxial stacks, including distributed Bragg reflector layers~\cite{He2017,Liu2017a}. In addition, the ability to access the far-field images of single QDs provides the possibility of determination of the emitter's dipole orientation, which can be used to align the orientations between the dipole and the local electric field for achieving the maximal light matter interaction~\cite{Bohmer2003,Takashima2020}. The determination of the dipole orientation is particularly important for certain applications.  One such application is the coupling of one of the split neutral exciton states to the polarized cavity mode of an asymmetric microresonator, to break the limit of 50~\% brightness in resonance fluorescence under the cross-polarization configuration~\cite{Wang2019a,Tomm2021}.
One critical aspect of the QD imaging approach is calibration of the optical images. In scanning techniques, calibrated encoders on sample translation stages (which in the case of electron microscopes, are sometimes inteferometrically referenced to stabilized lasers) are used for this purpose. In PL imaging, the alignment mark separation, whose value is known to within the accuracy of nanofabrication (which, for techniques like E-beam lithography, may come down to the accuracy of an interferometrically-referenced positioning stage), can provide a calibration reference. However, simple linear calibration based on the number of pixels that comprise the alignment mark separation inherently assumes an absence of distortion in the image. More sophisticated image calibration techniques that account for the presence of such distortion can be used to improve the accuracy and precision of camera-based PL imaging~\cite{Copeland2018}.

\section{\textit{In situ} lithography based positioning technique}

In both scanning-based and PL imaging-based positioning techniques, QD spatial information is extracted in one setup and definition of photonic nanostructures is implemented separately in another tool, typically an E-beam lithography system. These approaches require fabrication of alignment marks, location of those alignment marks and QDs, transferring the coordinates between the inspection system and the fabrication system, and fabrication using those transferred coordinates and the alignment mark location technique inherent to the fabrication system. This use of separate systems for identifying the QD locations and lithographically defining the nanostructures that surround them can potentially introduce errors whose origin can be hard to resolve. Thus, textit{in situ} lithography processes were developed to realize the extraction of QD positions and the definition of the photonic nanostructures in the same setup, with both optical and E-beam lithography approaches having been demonstrated. Amongst the many challenges addressed by these works is that of lithography within a cryogenic environment\cite{Lee2006}.

\subsection{\textit{In situ} photolithography-based positioning}

The first \textit{in situ} positioning technique for creating nanophotonic devices with single located QDs used far-field optical lithography to create QD-micropillar structures with controlled light-mattering interactions in both the weak and strong coupling regimes~\cite{Dousse2008}. In this seminal work, QDs in a thin GaAs layer sandwiched in between of the two sets of Al$_{0.1}$Ga$_{0.9}$As/Al$_{0.95}$Ga$_{0.05}$As DBR mirrors (thereby forming a planar cavity) were excited by using a red laser whose energy was higher than the bandgap of GaAs. By repeatedly scanning the samples via a set of closed-loop piezo steppers, the positions of QDs are extracted by fitting the QD intensity distribution with a Gaussian profile – this is essentially the same approach as subsequently used by Thon \textit{et al.}\cite{Thon2009} as described earlier. However, in contrast to that work, lithography is subsequently performed \textit{in situ}. Moving the targeted QD to the position of the excitation laser spot, another green laser was then introduced through the same optical path to expose the photoresist that was earlier spun on the surface of the sample, as shown in \textbf{Figure  \ref{fig.3}(a)}. The exposed resist serves as a circular mask for defining the micro-pillar cavities (which are subsequently etched through the semiconductor layers). The resonance wavelength of the micro-pillar cavity mode is determined by the radius of the exposed area, which can be controlled by the laser intensity. So far, the \textit{in situ} photolithography technique has most notably been used to deliver high-performance micro-pillar cavities with deterministically embedded QDs~\cite{Nowak2014a} (see  \textbf{Figure  \ref{fig.3}(b,c)}, and ultimately, the limited resolution of the laser lithography approach (in comparison to E-beam lithography or a deep-UV stepper) is best suited for such relatively straightforward mask geometries. More recently, this approach has been used as part of a two-step lithography process~\cite{Kolatschek2019}, where \textit{in situ} lithography is used to define alignment marks whose separation with respect to the targeted QD is precisely known. These alignment marks are then used in the lithography of more complicated photonic nanostructures, e.g., so that circular Bragg grating cavities or photonic crystals can be fabricated through implementation of an additional E-beam lithography process.

\subsection{\textit{In situ} E-beam lithography-based positioning}

\begin{figure}
	\includegraphics[width=0.95\linewidth]{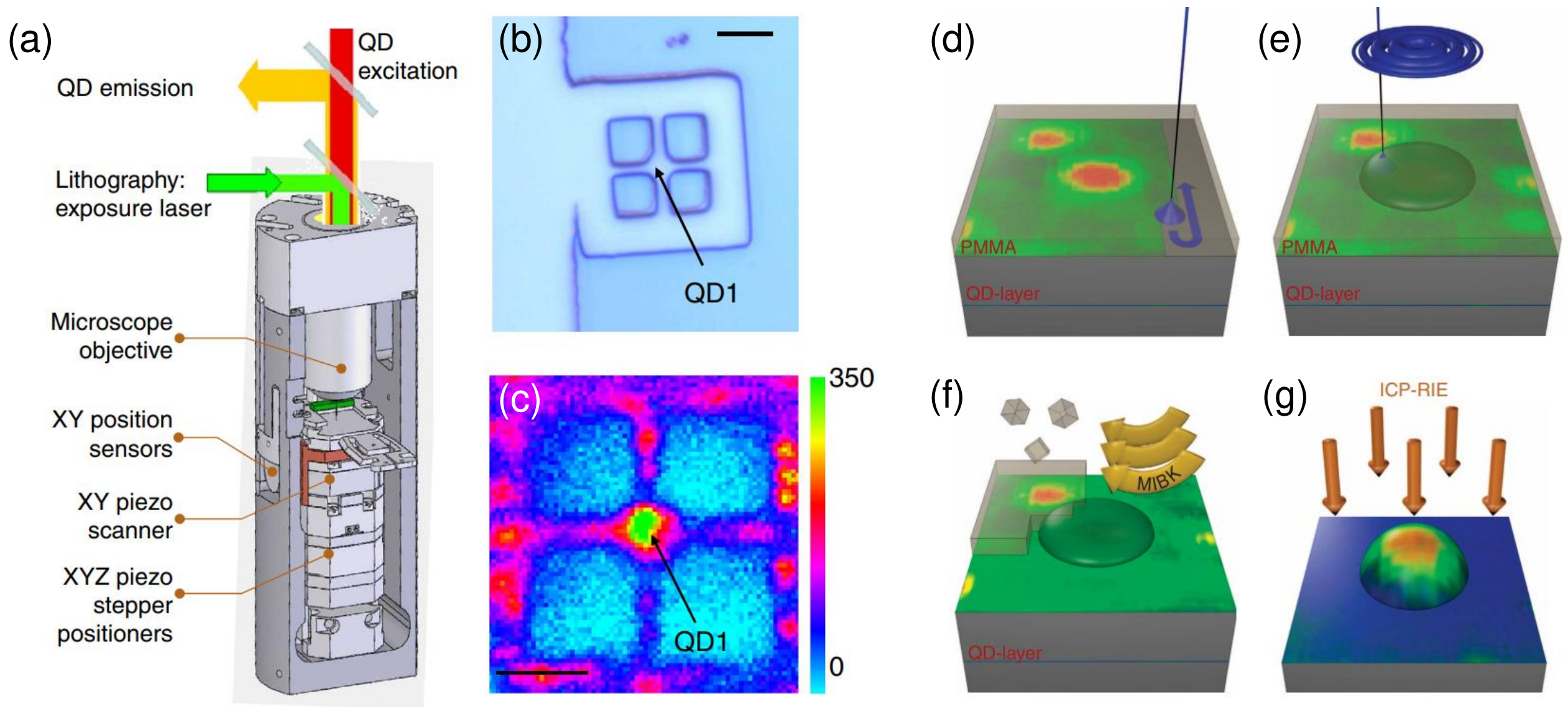}
	\centering
	\caption{In-situ lithography-based positioning technique. (a) In-situ optical lithography setup, (b) exposed photon resist and (c) fabricated QD-micropillar coupled device. (d-g) in-situ EBL process for deterministic QD-micro-lens device. ICP-RIE: inductively-coupled plasma reaction ion etch; MIBK: methyl isobutyl ketone. (a-c) Reproduced from~\cite{Nowak2014a}, \copyright 2014 Macmillan Publishers Ltd, with permission of Springer. (d-g) Adapted from~\cite{16gschrey2015highly}, \copyright 2015 Macmillan Publishers Ltd, with permission of Springer.}
	\label{fig.3}
\end{figure}

QD location with \textit{in situ} E-beam lithography has been pioneered by the Reitzenstein group in TU Berlin~\cite{Gschrey2015,16gschrey2015highly}, and very recently summarized in a comprehensive review~\cite{Rodt2021}. They have combined a low-temperature cathodoluminescence (CL) spectroscopy setup for locating single QDs together with a customized SEM in which the scan coils can be controlled for E-beam lithography. Similar to the aforementioned \textit{in situ} photolithography, E-beam resist was coated on the surface of the sample before cooling it down using a cryogenic insert within the custom SEM. Instead of optically pumping the QDs, the electron beam excites the CL from the QDs. By carefully choosing the dose of the scanning E-beam, a CL map of QDs over a field of view of $\approx$20 $\mu$m$\times $30 $\mu$m was obtained without significantly exposing the resist, see \textbf{Figure  \ref{fig.3}(d)}. The positions of individual QDs were extracted by fitting the 2D intensity distributions of the CL. With the extracted QD positions known, an E-beam lithography process was then implemented in the same system to define the photonic nanostructure mask around the QD, without moving the sample at all, as shown in \textbf{Figure  \ref{fig.3}(e)}. After warming up, the photonic nanostructures are fabricated by using a dry etching process to transfer the mask pattern into the semiconductor layer, see \textbf{Figure  \ref{fig.3}(f,g)}. The unique combination of E-beam lithography and CL is very promising for site-selective exposure of photonic nanostructure with respective to the positions of quantum emitters. The first proof-of-concept demonstration was the realized with simple mesa structures~\cite{Gschrey2013}. Later micro-lenses with deterministically coupled QDs were successfully fabricated via a specially engineered E-beam exposure process, resulting in a highly indistinguishable single-photon source~\cite{16gschrey2015highly}. The potential of \textit{in situ} E-beam lithography was recently further demonstrated by integrating QDs with on-chip nanophotonic multi-mode interferometer~\cite{Schnauber2018} (MMI) and with waveguides that were heterogeneously integrated with silicon nitride photonic circuits~\cite{Schnauber2019}, showing great potential for creating complicated and large-scale quantum photonic circuits for on-chip quantum simulation/computation.

\section{Summary of positioning approaches}

Table 1 lists several figure of merit for the different positioning techniques discussed in this review. Each method possesses its own advantages and disadvantages. In general, scanning techniques have longer images acquisition times, and those based on non-optical microscopy modalities obviously do not provide optical information (e.g. wavelength and linewidth) about the QDs that is generally needed for subsequent nanophotonic device fabrication. In addition, the QD positions extracted from AFM and SEM images may not be coincident with the center of the electric dipoles in the QDs. On the other hand, these techniques can take advantage of well-calibrated translation stages with encoders that ensure an accurate recording of position during the scan. The PL imaging method is competitive with a fast data acquisition speed, high positioning accuracy, optical properties available and an ability to simultaneously position multiple QDs in a large area. However, it requires careful calibration of the images to account for potential distortions, and like the scanning methods, requires that the coordinates determined from the positioning system be faithfully mapped to the system used for device fabrication, which could introduce errors. In contrast, the \textit{in situ} optical lithography technique does not require the separate fabrication of alignment marks, but on its own is limited to specific geometries such as micropillars or lenses. However, it can also be used to create alignment marks that are then utilized in E-beam lithography, and it could foreseeably be combined with photoluminescence imaging to increase throughput. Finally, \textit{in situ} E-beam lithography with cathodoluminescence-based QD location has many potential advantages in terms of fidelity in directly using extracted QD positions in fabrication, and the ability to create complex patterns. However, it requires the most significant amount of specialized equipment of all the techniques described.

\begin{table}
	\caption{ Comparison of the figures of merit for nanoscale positioning techniques.}
	\begin{tabular}[htbp]{@{}p{2.5cm}p{2.0cm}p{2cm}p{2.0cm}p{2cm}p{2.0cm}p{2.0cm}p{2.0cm}@{}}
		\hline
		                     & Data acquisition time & Positioning accuracy  &  Surface characterization& Arbitrary structures & Positioning multiple QDs & Alignment marks required & optical information provided \\
		\hline
		SEM  &$\approx$1 min  & $\approx$10 nm~\cite{Kuruma2016} & Required & Yes &Yes& Yes & No \\
		AFM  & Few minutes  & $\approx$30 nm~\cite{Hennessy2007} & Required & Yes & Yes & Yes & No\\
		PL-scan  & Few minutes  & $\approx$10 nm~\cite{Thon2009} & Not required & Yes & No & Yes & Yes\\
		PL-imaging  & 1 s  & $\approx$5 nm~\cite{Liu2017} & Not required & Yes & Yes & Yes & Yes\\
		\textit{In situ} photolithography  & Few minutes  & $\approx$50 nm~\cite{Dousse2008} & Not required & No & No & No & Yes\\
		\textit{In situ} Ebeam-thography  & Few minutes  & $\approx$30 nm~\cite{Gschrey2015} & Not required & Yes & Yes & No & Yes\\
		\hline
	\end{tabular}
\end{table}

\section{Applications of positioning techniques in quantum photonics}

\subsection{Quantum light sources in the weak coupling regime}
\begin{figure}[h]
	\includegraphics[width=0.95\linewidth]{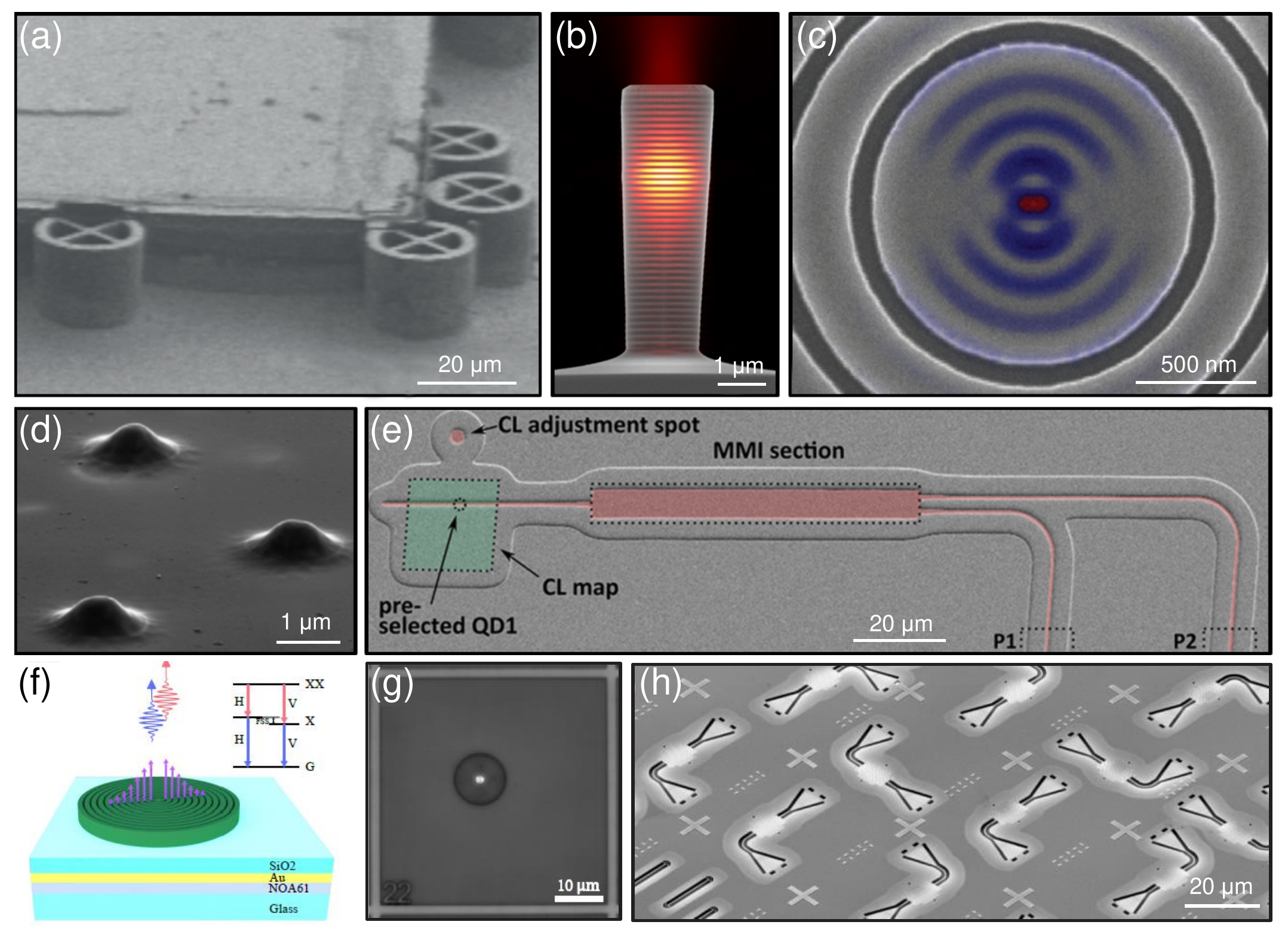}
	\centering
	\caption{Deterministically fabricated weakly-coupled system and propagating modes-coupled device based on QDs. (a),(b)Micro-pillar, (c) Suspended Bulleye cavity, (d) Micro-lens, (e) Waveguide, (f-g) Circular Bragg resonator on highly-efficient broadband reflector (CBR-HBR) device, (h) Photonic crystal waveguide. (a) Reproduced from \cite{Senellart2017}, Copyright \copyright 2017 Macmillan Publishers Ltd. With permission of Springer. (b) from \cite{Liu2017a}, with permission of Springer. (c) Reprinted from  \cite{Sapienza2015}. Copyright \copyright 2015 Macmillan Publishers Ltd. With permission of Springer. (d) from \cite{16gschrey2015highly}. Copyright \copyright 2015 Macmillan Publishers Ltd. With permission of Springer. (e) Reproduced with permission from~\cite{Schnauber2018} Copyright \copyright2018 American Chemical Society. (f-g) from~\cite{Liu2019}, Copyright \copyright  2019, The Author(s), under exclusive licence to Springer Nature Limited. With permission of Springer. (h) Reprinted from \cite{Pregnolato2020}, with the permission of AIP Publishing.}
	\label{fig.4}
\end{figure}

For specific applications such as high-performance quantum light sources, one of the challenges is to efficiently out-couple the emitted photons to the collection optics. Moreover, one may want to take advantage of cavity QED effects, such as Purcell enhancement, which increase the radiative rate. Such tasks can be fulfilled by employing QD-cavity systems operating in the weak coupling regime or coupling the emission from QDs into propagating optical modes in tailored waveguides~\cite{Sapienza2015,Pregnolato2020,Liu2019,Liu2017a,Su2018,16gschrey2015highly,Somaschi2016Near}(for example, slow light photonic crystals), as summarized in \textbf{Figure  \ref{fig.4}}. In such cases, the precise location of the emitter within the electromagnetic field plays an important role.
Both \textit{in situ} lithography and PL imaging have been shown to be reliable tools to create weakly coupled QD-cavity system. By deterministically fabricating micropillars with embedded QDs and carefully choosing the pillar diameters, most of the emitted single-photons are funneled into the cavity mode and subsequently collected by a microscope objective in free-space, resulting in a very high brightness of the single-photon source~\cite{He2017,Liu2017a,Somaschi2016Near,Wang2019}(defined here in terms of the probability that an excitation pulse results in a collected single photon). The pronounced Purcell effect provided by interaction with the cavity mode effectively reduces the pure dephasing process due to the solid-state environment and thus offers coherent single-photon emission (the shortening of the QD radiative lifetime limits the extent to which dephasing processes can influence the coherence of the emission). The state-of-the-art QD micropillar single-photon sources exhibit simultaneous high-degrees of brightness, single-photon purity and photon indistinguishability, which significantly outperforms their spontaneous parametric downconversion (SPDC) counterpart and therefore open unprecedented opportunities in quantum simulation~\cite{Senellart2017}. Besides single-photon sources, entangled photon pairs are also a crucial resource in many quantum information processing protocols based on quantum teleportation and entanglement swapping. Two polarization entangled photons with slightly different colors can be generated in a triggered manner by utilizing the bi-exciton(XX) exciton (X) cascaded decay process in QDs. To boost the brightness of the QD entangled sources, one can either couple the X and XX states to two different high-Q cavity modes whose frequency separation matches the X-XX separation, or to one low-Q cavity with broad bandwidth. The former was demonstrated in a micropillar molecule via \textit{in situ} photolithography~\cite{Dousse2010b} while the later were recently implemented in a circular Bragg resonator by using PL imaging~\cite{Liu2019}, with both showing significantly enhanced brightness and high-degree of entanglement fidelity. For on-chip quantum information processing, it is more desirable to couple the QDs to propagating waveguide modes that can further route the emitted single-photons in-plane for quantum interferences. Near-unity coupling efficiency between the QDs and photonic waveguides has been achieved by probabilistic coupling~\cite{Kirsanske2017} and very recently by deterministic coupling via PL imaging technique~\cite{Pregnolato2020}.

\subsection{Strongly coupled systems}

\begin{figure}[h]
	\includegraphics[width=0.95\linewidth]{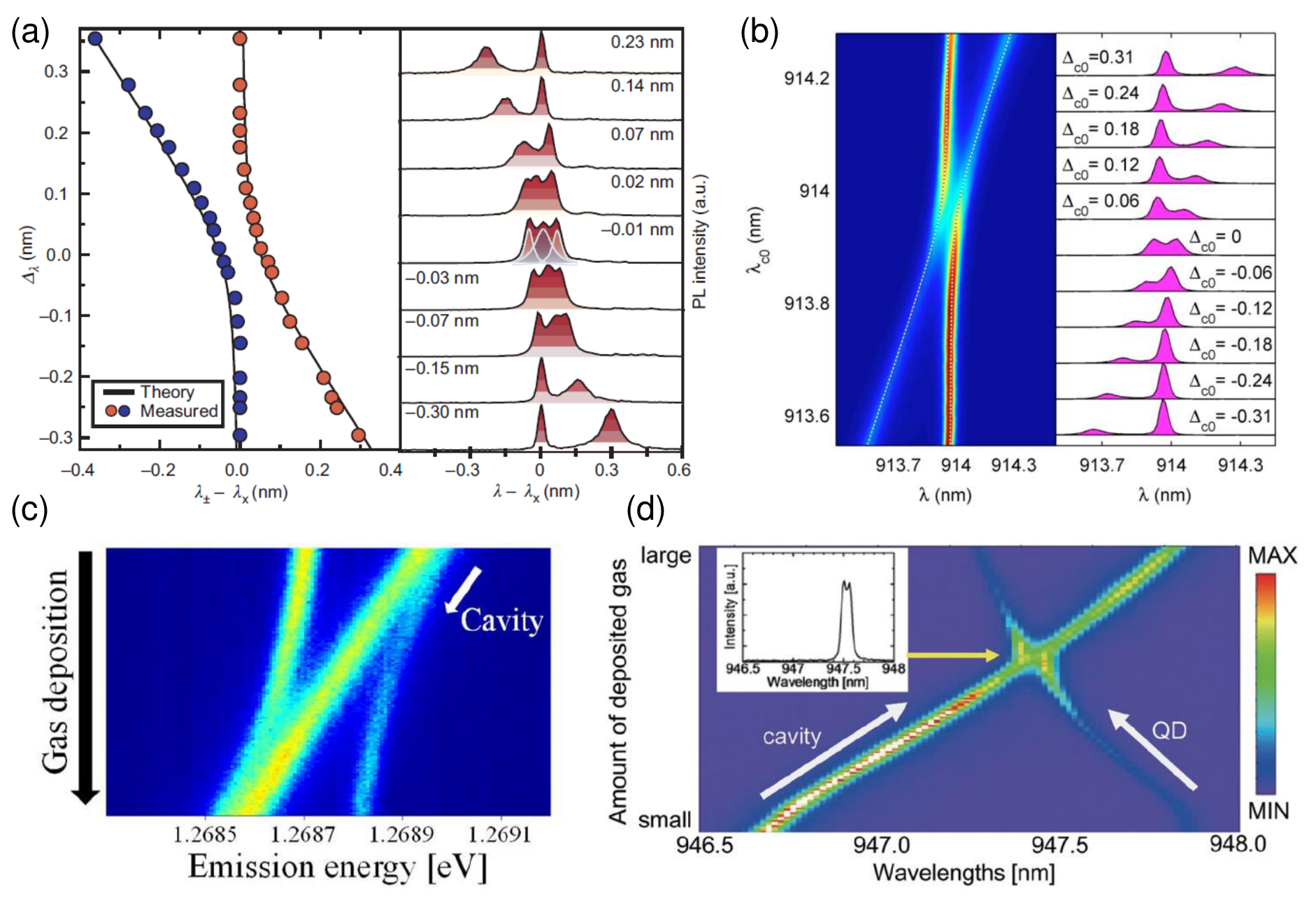}
	\centering
	\caption{Deterministically fabricated strongly-coupled system based on QDs located in L3 photonic crystal cavity devices, consisting of three missing air holes within a hexagonal lattice of air holes in a thin dielectric slab. (a) Reproduced from \cite{Hennessy2007} Copyright \copyright  2007 Nature Publishing Group. With permission of Springer. (b) from \cite{Thon2009}, Copyright \copyright 2009 American Institute of Physics. With the permission of AIP publishing. (c) Reprinted from \cite{Kojima2013Accurate}, Copyright \copyright 2013 American Institute of Physics. With the permission of AIP publishing. (d) from \cite{Kuruma2016}, with the permission of AIP publishing.}
	\label{fig.5}
\end{figure}

The nature of light-matter interaction changes dramatically when moving from the weak coupling regime of perturbative interactions to the strong coupling regime in which a new quantum superposition state (an ‘atom-photon molecule’) that is half-light and half-matter is formed. The strongly coupled QD-cavity system in the solid-state is particular appealing due to the ability to realize important quantum information tasks such as quantum logic gates and entanglement of indistinguishable quantum systems~\cite{Gouet2012}. In order to reach the strong coupling regime, single QDs must be accurately placed at very specific positions within the high-Q cavities where intensity of the electromagnetic field is maximal. Despite the fact that the realization of strong coupling is more technologically challenging, strong coupled QD-cavity systems (examples are shown in \textbf{Figure  \ref{fig.5}}) have been readily accessed by using a variety of the positioning techniques developed along the years as shown in this review~\cite{Hennessy2007,Thon2009,Kojima2013Accurate,Kuruma2016}, indicative of the power of these nanoscales positioning techniques.

\subsection{Identifying the influence of nanofabrication on the optical properties of QDs}

The application of QDs in quantum photonics, such as quantum light sources and strongly coupled QD-cavity systems, involves the fabrication of nanoscale photonic structures, which inevitably bring the QDs in proximity to the surfaces formed by dry etching. It is well-known that the defect states on the surfaces could potentially lead to a carrier tunneling and unstable charge environment for the QDs, resulting in blinking and spectral diffusion. However, it is still an open question as to what extent nanofabrication influences the optical properties of the QDs embedded in photonic nanostructures, and how large a separation between the QDs and these surfaces is needed. To answer this question, one needs to accurately control the distance between the QDs and the etched surfaces and characterize the QDs before and after the dry etching. As shown in \textbf{Figure  \ref{fig.6}(a-c)}, by employing the PL-imaging technique to place dry etched surfaces closed to the QDs with finely controlled distances~\cite{Liu2018}, the authors demonstrated that QD linewidths are significantly broadened but changes of radiative efficiency are negligible when this separation is within a couple hundred nanometers, see \textbf{Figure  \ref{fig.6}(d)}. Furthermore, atomic layer deposition was shown to be effective in stabilizing the spectral diffusion and partially recovering the QD linewidth. With aid from nanoscale positioning tools, more efforts are expected to help understand how to recover the Fourier transform limited single-photons from QDs broadened by the etched surfaces, e.g., exploring surface passivation and oxide encapsulation \cite{Manna2020} to improve optical properties of QDs and using gated structures to control the local charge environment~\cite{Zhai2020,Uppu2020}.

\subsection{Correlating surface morphology and QD locations}

\begin{figure}[h]
	\includegraphics[width=\linewidth]{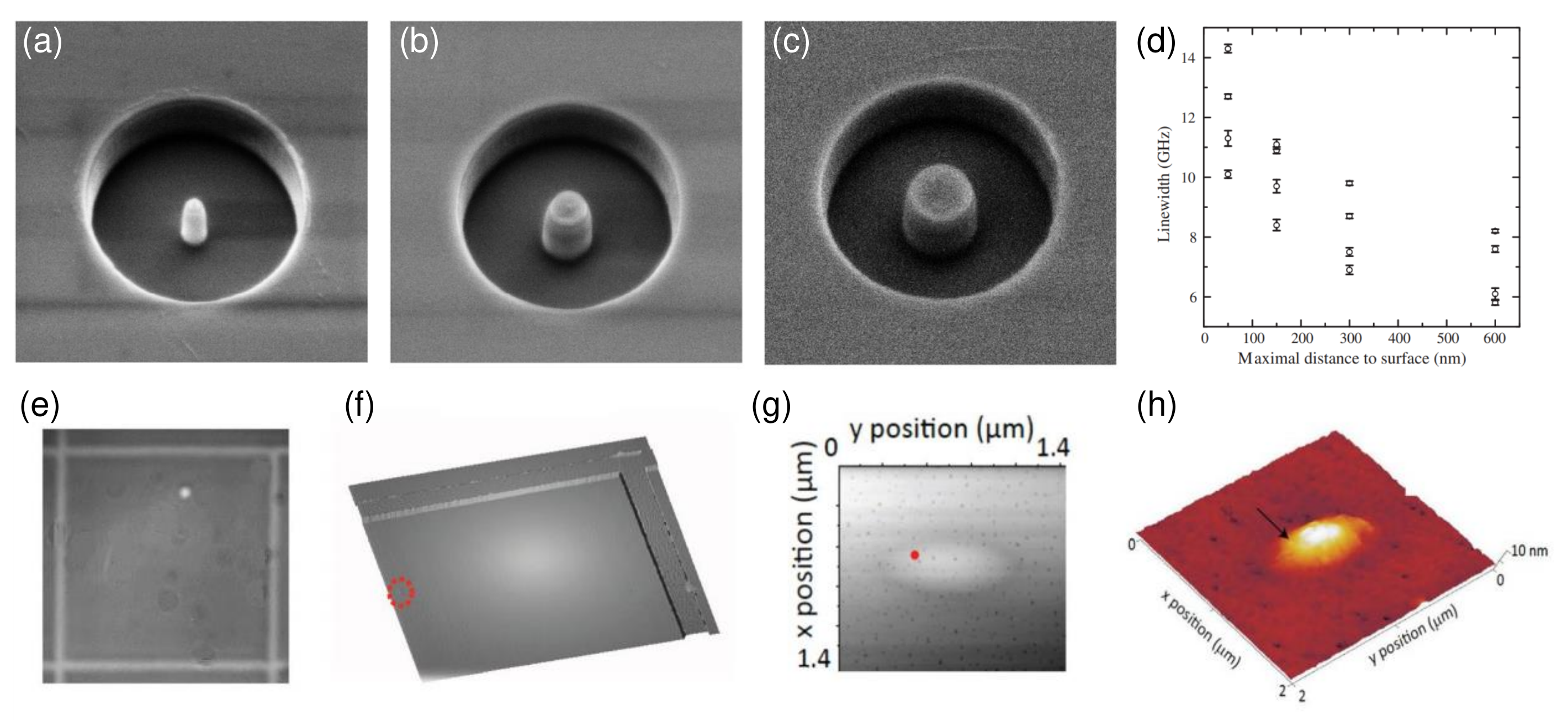}
	\caption{Other applications of positioning technique for identifying the influence of nanofabrication on the optical properties of QDs (a-d), and correlating surface morphology and QD locations (e-h). (a-d) Reproduced from \cite{Liu2018}, Copyright \copyright 2018 American Physical Society. (e-h) Reprinted from \cite{Sapienza2017}. \href{https://creativecommons.org/licenses/by/4.0/}{CC BY 4.0}.}
	\label{fig.6}
\end{figure}

One of the potential drawbacks of the AFM and SEM-based positioning techniques is that some specific surface morphologies may not necessarily be associated to the embedded QDs. To this end, the authors of a recent study fabricated alignment marks on the surfaces of a few different MBE-grown QD wafers grown by different collaborators and took AFM images in locations for which PL images and spectroscopy indicate the presence of single QDs~\cite{Sapienza2017}, as shown in \textbf{Figure  \ref{fig.6}(e,f)}. The authors found that the small bumps shown in ref. \cite{Kuruma2016} are not the universal features of the embedded QDs in all the investigated samples. They also observed other surface features, e.g., deep holes were seen above the buried QDs by using AFM, with the specific dimensions dependent on the growth. In addition, for samples showing the small bumps, the authors overlaid the PL image and the AFM images for one QD under the same coordinate system defined by the alignment mark. They found that the center of the dipole emission in the QD was at the periphery of the surface bump, see \textbf{Figure  \ref{fig.6}(g,h)}, indicating that the positioning technique based on surface characterizations may induce offsets between the QD and the maximal field of the cavity mode, and consequently reduce the achievable coupling strength.

\section{Conclusion}

The ability to precisely locate solid-state quantum emitters so that their position in photonic nanostructures is tightly controlled provides a powerful tool in fundamental research by controlling the strength of light-matter interactions, e.g., probing the local density of optical states of photonic nanostructures with embedded QDs~\cite{Wang2011a}. At the device level, electrically pumped quantum light sources are highly desirable in real applications and deterministically fabricated single-photon LEDs were recently demonstrated with \textit{in situ} photolithography~\cite{Sartison2019a}. Eventually, electrically pumped coherent single-photons may be achieved by resonantly exciting QDs with micro-lasers on the same chip. To this end, the device yields will be appreciably improved with the deterministically fabricated devices. The positioning technique may also play an essential role in filter-free resonance fluorescence by pumping multiple QDs via an on-chip waveguide and collecting single-photons from the top~\cite{Huber2020}. With the positioning tools described in this review in hand, there is the opportunity to scale to multiple QD devices to build quantum networks with complexity and functionality far beyond single QDs devices based on probabilistic coupling. Along this line, Hong-Ou-Mandel interference from two remote QDs was demonstrated with both deterministically coupled micropillars~\cite{Giesz2015} and microlenses~\cite{Thoma2017}. Moving forward, spin-photon entanglement~\cite{Gao2012}, spin-spin entanglement~\cite{Stockill2017} and entanglement swapping~\cite{BassoBasset2019,Zopf2019} will greatly benefit from the positioning techniques. Last but not the least, although the positioning techniques covered by this review were developed in the context of epitaxial QD devices, most of these techniques can be easily adopted to other solid-state quantum emitters, e.g., color centers in diamond~\cite{Su2009,Aharonovich2011} and SiC~\cite{Castelletto2013}, and defect states in GaN~\cite{Berhane2017,Zhou2018} and 2D materials~\cite{Tran2016,He2015}. The significantly improved device yield and the controlled light-matter interactions enabled by nanoscale positioning of solid-quantum emitters may provide unprecedented opportunities in advancing the active optoelectronic devices and integrated quantum photonic technologies.

\section{Acknowledgements}
The authors thank Serkan Ates and Ashish Chanana for helpful feedback, and Anshuman Singh for his early contributions to developing the structure of this review. This research was supported by National Key R\&D Program of China (2018YFA0306100, 2018YFA0306302), the National Natural Science Foundation of China (11874437, 61935009), Guangzhou Science and Technology Project (201805010004), the Natural Science Foundation of Guangdong (2018B030311027), and the national super-computer center in Guangzhou.

\section{Conflict of Interest}
The authors declare no conflict of interest.
\medskip

\medskip

%



\begin{figure}[h]
	\includegraphics[width=4cm]{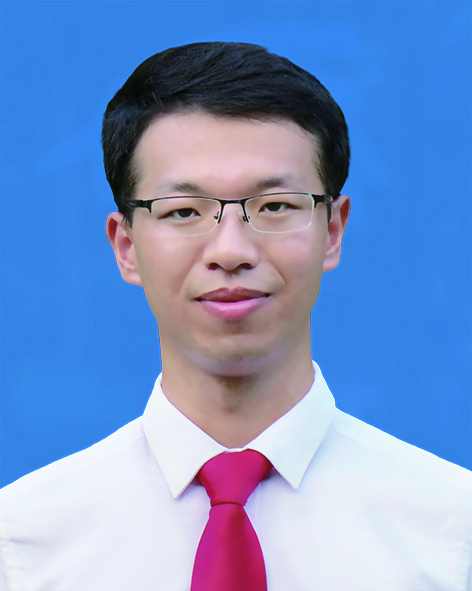}
	\caption*{Shunfa Liu is a Ph.D. student in School of Physics, Sun Yat-sen University. His current research interest includes quantum dot single-photon sources and nanofabrication.}
\end{figure}

\begin{figure}[h]
  \includegraphics[width=4cm]{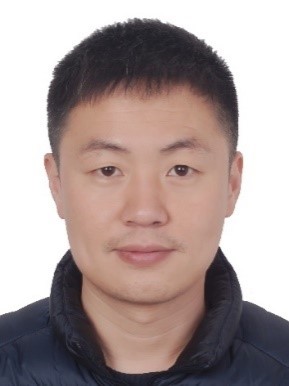}
  \caption*{Jin Liu obtained his Ph.D. degree in 2012 in Technical University of Denmark. He is currently a Professor in
  	School of Physics, Sun Yat-sen University, Guangzhou,
  	China. His research activities cover integrated optics,
  	solid-state quantum photonics and semiconductor
  	nanofabrication.}
\end{figure}

\begin{figure}[h]
  \includegraphics[width=4cm]{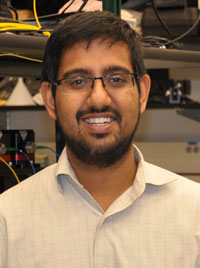}
  \caption*{Kartik Srinivasan obtained his Ph.D. in 2006 from the California Institute of Technology.  He is currently a Fellow at the National Institute of Standards and Technology (NIST) and the NIST/University of Maryland Joint Quantum Institute, where he is also an Adjunct Professor of Physics. His research is in the field of nanophotonics, with a focus on topics in photonic quantum information science, nonlinear optics, and metrology.}
\end{figure}


\end{document}